\documentclass[12pt]{article} 
\usepackage[hyperfootnotes=false]{hyperref}
\usepackage{amsmath}
\usepackage{amssymb}
\usepackage{graphicx}
\usepackage{xcolor}
\newlength{\abstractwidth}
\usepackage{ wasysym } 
\usepackage{mciteplus} 
\usepackage{subcaption}
\usepackage{bbold}
\usepackage{slashed}

\newcommand{\be}{\begin{equation}}
\newcommand{\bea}{\begin{eqnarray}}
\newcommand{\eea}{\end{eqnarray}}
\newcommand{\beq}{\begin{equation}}
\newcommand{\ee}{\end{equation}}
\newcommand{\eeq}{\end{equation}}

\newcommand{\half}{{1\over 2}}
\newcommand{\stimes}{{\scriptstyle \times}}

\def\la{\label}

\def\32{{3 \over 2 } }

\def\ba{\begin{eqnarray}}
\def\ea{\end{eqnarray}}

\def\simleq{\; \raise0.3ex\hbox{$<$\kern-0.75em
      \raise-1.1ex\hbox{$\sim$}}\; }
\def\simgeq{\; \raise0.3ex\hbox{$>$\kern-0.75em
      \raise-1.1ex\hbox{$\sim$}}\; }

\def\nref#1{(\ref{#1})}
\def \la {\label}   
 
\def \half {{1\over 2}}	
\def\nref#1{(\ref{#1})}

\def \be {\beta}

\def \Ac {\mathcal{A}}

\def \beq { \begin{equation}}
\def \eeq {\end{equation}}

\def \l {\left(}
\def \r {\right)}

\newenvironment{longeq}
{\begin{equation}\aligned}
{\endaligned\end{equation}}

\def \beqs{\begin{longeq}}
\def \eeqs{\end{longeq}}

\renewcommand \kappa {\varkappa}

\begin{document}

\begin{titlepage}
  \bigskip

  \bigskip\bigskip

  \bigskip

\begin{center}
 
\centerline
{\Large \bf {Humanly traversable wormholes}}
 \bigskip

 \bigskip
{\Large \bf { }} 
    \bigskip
\bigskip
\end{center}

  \begin{center}

 \bf {Juan Maldacena$^1$ and  Alexey Milekhin$^{2}$    }
  \bigskip \rm
  
\bigskip
 $^1$Institute for Advanced Study,  Princeton, NJ 08540, U.S.A.  

 \rm 
 \bigskip
   $^2$Physics Department, Princeton University, Princeton, NJ 08544, U.S.A.\\
\rm
 \bigskip

  \bigskip \rm
\bigskip
 
\rm

\bigskip
\bigskip

  \end{center}

 \bigskip\bigskip
  \begin{abstract}
We point out that there can be humanly traversable wormhole solutions in some previously considered theories for physics beyond the Standard Model,
namely the Randall–-Sundrum model.

 \medskip
  \noindent
  \end{abstract}
\bigskip \bigskip \bigskip

\vspace{1cm}

\vspace{2cm}

  \end{titlepage}

   \tableofcontents


\section{Introduction } 
  
   Traversable wormholes are a staple of the science fiction literature. 
   In classical general relativity, they are forbidden by the average null energy condition \cite{Galloway:1999bp,Galloway:1999br,Witten:2019qhl}. Interestingly, they are allowed in the quantum theory, but with one catch, the time it takes to go through the wormhole should be longer than the time it takes to travel between the two mouths on the outside\footnote{  This also implies that they can not be converted into time machines \cite{Morris:1988tu}.   }. Nevertheless they are interesting configurations that are allowed by the laws of physics as we know them. In fact,  based on the initial construction in \cite{Gao:2016bin}, four dimensional traversable wormholes were constructed in 
 \cite{Maldacena:2018gjk}, see also \cite{Fu:2019vco,Fu:2018oaq}. The construction in \cite{Maldacena:2018gjk} 
involved ingredients that are present in the Standard Model, but only at short distances. Therefore only microscopic wormholes were found. 

In this paper, we revisit the question and we engage in some  ``science fiction''. Namely, we will introduce a dark sector with desirable properties for constructing    macroscopic traversable wormholes. This dark sector only interacts via gravity with the Standard Model. Our main point is to emphasize the consistency of such wormholes with the laws of physics as we presently know them, using only previous ideas for physics beyond the Standard Model. In particular, we will use a version of the Randall-Sundrum model \cite{Randall_1999}.

Our construction needs  a dark sector consisting of a four dimensional conformal field theory with a U(1) symmetry that is gauged by a four dimensional (dark) gauge field. A simple example would be a theory of many massless fermions coupled to a $U(1)$ gauge field \cite{Maldacena:2018gjk}. Better wormholes are possible by using a  Randall Sundrum II model  
 \cite{Randall:1999vf} with a $U(1)$ gauge field.
%
This model allows for large enough wormholes that could be traversed humanely, i.e.  surviving the tidal forces.   Using them,  one could travel in less than a second between distant points in our galaxy. A second for the observer that goes through the wormhole. It would be tens of thousands of years for somebody looking from the outside. 

This paper is organized as follows. 
In section two,  we review the main idea for the construction of traversable wormholes and discuss the example involving massless charged fermions. 
In section three,  we consider the Randall Sundrum II model and argue that we can have wormholes large enough to allow for human travelers. 
 In the discussion section we mention some further practical issues that make such wormholes  problematic in practice.

     \vskip .5 cm 
 
\section{Review of traversable wormholes}

\subsection{Preliminaries } 

The theory we will be considering consists of gravity, plus a $U(1)$ gauge field and a four dimensional matter theory with a $U(1)$ symmetry which we are gauging with the dynamical gauge field
\beq \la{theor} 
S = { 1 \over 16 \pi G_4 } \int d^4 x \sqrt{g} R - { 1 \over 4 g^2_4 } \int d^4 x \sqrt{g} F_{\mu \nu} F^{\mu \nu}+  S_{\rm matt}[g_{\mu \nu} , A_\mu ]  
\eeq
where $F_{\mu \nu}$ is the usual field strength, $F = dA$. Both $A_\mu$ and the matter theory are part of a dark sector. 

In \cite{Maldacena:2018gjk},  the  matter theory, $S_{\rm matt}$,   was given by $N_f$ free massless fermions charged under the $U(1)$. 
Here we consider a matter theory which is holographic, namely it has an $AdS_5$ dual described by five dimensional gravity and a five dimensional $U(1)$ gauge field which we also call $A_\mu$. We will later discuss it in more detail.

A crucial property for the construction  is the following. In the presence of a constant magnetic field, the four dimensional matter theory flows to a two dimensional theory with a central charge that scales linearly with the total magnetic flux, 
\beq \label{ctwo} 
c_2 \propto B \Ac = 2 \pi q 
\eeq 
 where $B$ is the magnetic field,  $\Ac$ is the transverse area and $q$ the integer magnetic flux. 
We can imagine dividing the transverse area into flux quanta and we have a certain  central charge per flux quantum. For the case 
with $N_f$ free fermions, the central charge was  
$c_2 = N_f q$, due to $q$ 2d massless modes \cite{Callan:1982ac, Rubakov:1982fp}.
The transition between the four dimensional description and the two dimensional one happens at a distance scale 
\beq \la{d24}
l_B \propto { 1 \over \sqrt{B}}
\eeq 

\subsection{Magnetic black holes } 

\begin{figure}[!ht]
~~~~~~~ ~~~\includegraphics[width =15cm,height=7cm]{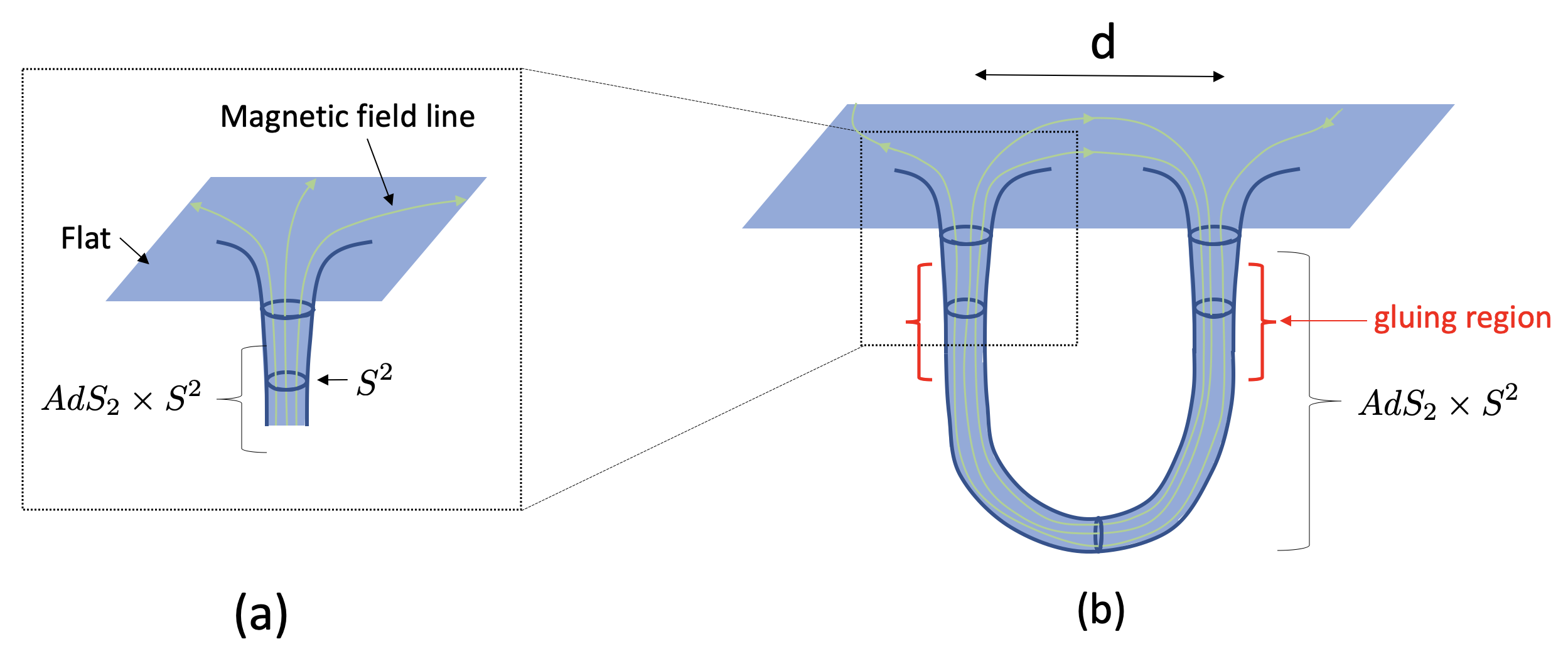}
 \caption{ (a) We see a sketch of the geometry of a single extremal black hole. It develops a long throat with a locally $AdS_2 \times S^2$ geometry. (b) The wormhole geometry. We start with two separate throats and we connect them.  The geometry of the connected throats is again $AdS_2 \times S^2 $. The two mouths of the wormhole are at some distance $d$. The red brackets  indicates the region of overlapping validity of the metrics 
 \nref{Metr} and \nref{GlobalAdS}. The green lines denote the magnetic field lines. They form closed circles, entering one mouth, traveling through the wormhole, exiting the other mouth and then going back to the first mouth in the ambient space.  
}
\label{fig:BHandWH}
\end{figure}

We start with the geometry of an extremal  magnetically charged black hole 
\begin{align}  
\label{Metr}
ds^2 &=  -f dt^2 + { dr^2 \over f }  + r^2 ( d\theta^2 + \sin^2 \theta d\phi^2 ) ~,~~~~~~~~~ A = { q \over 2 } \cos \theta d\phi   \\
f &= \left( 1-  \frac{r_e}{r} \right)^2 ~,~~~~~~r_e \equiv { \sqrt{\pi}     q  l_p \over g_4  }  ~,~~~~~~l_p \equiv \sqrt{G_4} ~,~~~~~~M_e = { r_e \over G_4} \label{redef}
\end{align}
where $q$ is the (integer) magnetic charge and $M_e$ the mass at extremality. $r_e$ sets the radius of curvature of the geometry in the 
near horizon region, and also the size of the two sphere.  At this extremal limit the geometry 
develops an infinite throat (as $r\to r_e$) where the redshift factor, $g_{tt}$, 
becomes very small but the size of the sphere remains constant, see Figure \ref{fig:BHandWH}(a). 

When we put the matter theory on  this background, it leads to a two dimensional CFT at scales less than \nref{d24}. 
Notice that the proper size of the magnetic field is 
\beq 
B   = { q \over 2 r^2 } 
\eeq
which for large $q$, $q\gg 1$, is much larger than the inverse size of the sphere. This means that we can ignore the curvature of the two sphere when we consider the transition between the four dimensional theory and the two dimensional one. For that reason we can study this transition in flat space. We will later discuss in detail how this occurs for a holographic theory. 

\subsection{The wormhole ansatz} 

We will now consider two black holes with opposite charges at some distance $d$ from each other. Naively, they would attract  and coalesce. However, let us first imagine they are somehow held fixed so that we can  ignore this issue for the moment. We will return to it later. 

If we had two separate extremal black holes each of them would have an infinite throat like the one depicted in Figure \ref{fig:BHandWH}(a). We now change the geometry of the throats by connecting them to each other so that the geometry of the deep throat region is given by 
\beq  
ds^2 = r_e^2 \left[ - (\rho^2 +1) d\tau^2 + { d \rho^2 \over (\rho^2 +1) } + ( d\theta^2 + \sin^2 \theta d\phi^2 ) \right]  ~,~~~~~~ - \rho_c \leq \rho \leq \rho_c \label{GlobalAdS}
\eeq
Notice that the size of the sphere is constant. This is a portion of the  $AdS_2 \times S^2$ geometry, which is  a solution of the equations of  gravity plus a magnetic field. 
Beyond  $\rho = \pm \rho_c$ the geometry is described by that of the extremal black hole  \nref{Metr}. The final geometry looks as in Figure \nref{fig:BHandWH}(b). It has a free parameter which is the total length of the wormhole. We will define this quantity as the ratio of the time in \nref{GlobalAdS} to the asymptotic time at infinity in \nref{Metr}  
\beq  \la{Ldef}
\ell = {t \over \tau } 
\eeq
In terms of $\ell$, 
 we can also find the relation between the radial coordinate in \nref{GlobalAdS} and 
\nref{Metr} 
\beq \rho = \ell { (r -r_e) \over r_e^2 } ~,~~~~~{\rm for } ~~\rho\gg 1 ~,~~~{\rm and }~~~~ r-r_e \ll r_e 
\eeq
The last two conditions ensures that we are far from the center of the wormhole but we are deep inside the throat. For $\ell \gg r_e$, there is an overlapping region where we glue the near horizon region  of the black hole \nref{Metr} to the wormhole region \nref{GlobalAdS}, see Figure \nref{fig:BHandWH}(b). 

After this step we do {\it not}  have a solution of the Einstein + Maxwell equations. In order to obtain  an actual solution we need to include the effects of matter. This matter behaves as a two dimensional conformal field theory with the central charge proportional to $q$, \nref{ctwo}. This two dimensional CFT can be viewed as composed by somewhat separate CFTs , each living  along a magnetic field line. These magnetic field lines form circles. They enter one mouth of the wormhole and they exit through the other, see Figure \nref{fig:BHandWH}(b). A two dimensional theory on a circle gives rise to a negative Casimir energy. This  gives some additional negative energy which turns the wormhole into a solution for a special value of the length parameter \nref{Ldef}.

We will not repeat here the full argument for the solution, which can be found
in detail in \cite{Maldacena:2018gjk}. This involves solving the Einstein
equations with a source given by the quantum stress tensor of the two
dimensional field theory \footnote{There are other contributions from the
matter stress tensor which are $SL(2)$ invariant in the throat region. Such
contributions can change slightly the black hole parameters. In particular,
they include contributions from the four dimensional conformal anomaly.  The
Casimir energy contribution from the effective two dimensional theory is the
leading contribution that violates the $SL(2)$ symmetry and it is the important
one for the construction of the wormhole.}. 

Imposing the Einstein equations,  we deform slightly the  metrics \nref{Metr},
\nref{GlobalAdS} and determine the parameter $\ell$ in \nref{Ldef}. We will
here only review a simple method to determine the final parameters of the
wormhole, which can be viewed as an energy minimization argument\footnote{ The
justification of this argument is that the geometry we described above is {\it
almost} a solution. We get a relatively small action for the variable $\ell$
from gravity and also one from the matter fields. So we can concentrate
extremizing the effective action for the variable $\ell$.}.

It turns out that the gravitational energy cost for joining the $AdS_2\times S^2$ region to flat space for ``length'' $\ell$ is the same as the energy above extremality for a black hole of temperature $T = 1/(2 \pi \ell)$ \cite{Maldacena:2018gjk}. This can sound more plausible if we notice that the rescaling  in \nref{Ldef} is the same as the rescaling we would have between Rindler time in the near horizon region of a near extremal black hole and the asymptotic time $t$. 
This energy  
 is twice the mass of a single near extremal black hole:
\beq
E_{\rm gravity} = 2M = 2 M_e + \frac{r_e^3}{G_4 \ell^2}  
\eeq
The two dimensional conformal field theory develops a negative Casimir energy equal to 
\beq 
E_{\rm Casimir} = - { c_2 \over 8 \ell } 
\eeq 
This also includes a contribution due to the conformal anomaly in $AdS_2$. 
Here $c_2$ is the total central charge of the two dimensional CFT \nref{d24}. 
In writing this expression we assumed 
that $d\ll \ell$ \footnote{The equations can also be solved when $d \sim \ell$. It is very similar, see 
\cite{Maldacena:2018gjk}.}.
Adding these two contribution and extremizing over $\ell$ we find 
\beq \la{Length}
\ell ={  16 r_e^3 \over G_4  c_2 }  
\eeq

Inserting \nref{Length} into the expression for the energy we find the binding energy 
\beq \label{Ebind}
E_{\rm bind} = E_{\rm gravity} + E_{\rm Casimir} - 2 M_e  =  -{ c_2 \over 16 \ell } = -  { c_2^2 G_4 \over 2^8  r_e^3 } 
\eeq
This means that forming the wormhole lowers the energy of the configuration, relative to that of two separate extremal black holes.  This energy is also important for the following reason. To keep the wormhole open we need the negative Casimir energy. If we send too much positive energy we will collapse the wormhole into separate non-extremal black holes. The maximum energy we can send, before that happens, is of order \nref{Ebind}. 

We can interpret this bound on the energy from the information transfer perspective. We can imagine sending signals through the wormhole 
using quanta with energy $\sim 1/\ell$. The constraint that the energy is smaller than $|E_{\rm bin}|$ in  (\ref{Ebind}) implies that the total number of quanta cannot exceed a number $\sim c_2$.
This means that we can not transfer more information than can be carried by the CFT through outside the wormhole. A similar bound exists for the
Gao--Jafferis--Wall protocol \cite{Maldacena:2017axo}. However, in our case we can talk about the information per unit time (namely of the order of $\sim c_2$ qubits 
per $\ell$), rather than total transferable amount.  

Let us note some geometric interpretations of $\ell$. 
 $\ell$ sets the redshift factor between the middle of the wormhole and the exterior. Namely, a massless particle with wavelength $1/r_e$ at the center of the wormhole would have an energy of order $1/\ell$ as seen from the outside. We can call this the energy gap of the wormhole, the minimum energy excitation of the configuration.  The binding energy \nref{Ebind} is larger by a factor of $c_2$.  In addition, $\ell$  sets the time it takes to travel through the wormholes as seen from the outside, which is $\pi \ell$. In deriving \nref{Length} we assumed that the distance $d$ between the two black holes is smaller than $\ell $, $d \ll \ell$. Even when we do not make that assumption we find that the time through the wormhole is always longer than through the outside, $\pi \ell > d$. On the other hand the proper traversal time for an observer going through the wormhole is of order $\pi r_e$, which is much shorter than $\ell$. 
 The ratio between these two times 
 \beq \label{Boost} 
 { \ell \over r_e} =  \gamma  \gg 1 
 \eeq 
    is the boost factor that a particle that starts non-relativistic outside would acquire at the center of the wormhole. 
 So, going through the wormhole is similar to accelerating to very high speeds and then decelerating. The difference is that the acceleration and deceleration is provided by gravity for free and one does not have to go through the ambient space. These wormholes are the ultimate roller coaster. 

We can now return to the problem of preventing the attraction of the two wormhole mouths. Notice that from the outside these have, up to a small correction \nref{Ebind},  the same mass and charge as two oppositely charged extremal black holes. We can make they orbit each other. It is possible to do this with an angular velocity $\Omega \ll 1/\ell $ that is smaller than the energy gap, so that we do not have a disruptive effect inside the wormhole. Of course, such a configuration would slowly emit gravity and dark-$U(1)$ waves which will eventually cause them to coalesce. In Appendix \ref{Rotation},  we explain that this happens at time scales that are parametrically longer than the traversal time, $\pi \ell$.

\subsection{Example with free fermions } 

In this subsection we review the values of above parameters for the case of $N_f$ free fermions. 
In that case $c_2 = N_f q$ where $q$ is the integer magnetic charge. 
This results in a wormhole length and energy gap 
\beq 
 { \ell \over r_e }  = 16 \pi { q \over g_4^2  N_f}     ~,~~~~~~~~E_{\rm bin} r_e = 2^{-8}  g_4^2 N_f^2  
 \eeq

This model is under control as long as $g^2_4 N_f < 1$. Otherwise the gauge field becomes strongly coupled.   
For reasons we will discuss in more detail later, due to tidal force reasons,  we need $r_e > 10^7 m$. If we further assume that the mass of the spaceship is about $10^3 kg$, we also need the condition $|E_{\rm bin}| > 10^3 kg$. This then implies that 
 \beq
 r_e |E_{\rm bin }| \propto N_f (g^2 N_f) > 10^7 m \times 10^3 kg  ~~~~~\to N_f > 10^{52} 
 \eeq 
 This value of $N_f$ is too large if we want the UV cutoff of the local field theory to be above the TeV scale, since we expect that $N_f < M_{pl}^2/({\rm Tev})^2 \sim 10^{32}$ due to Bekenstein-like arguments. Since this suggests that a strongly coupled theory might be desirable, it is natural to look for theories with an $AdS_5$ gravity description.

 \section{Traversable wormholes in the Randall-Sundrum II model } 
\label{FlowAdS}

In this section we will study a theory which includes a dark sector based on the Randall-Sundrum II model \cite{Randall:1999vf}.  The feature that is important 
for us is that it includes an $AdS_5$ region that extends towards the infrared. This can also be viewed as four dimensional CFT coupled to gravity. This point of view is useful, but not essential,  for describing the construction. It is useful because we can just show that this $AdS_5$ theory has the requisite general properties described in the previous section. The main property that we will use is the emergence of a two dimensional CFT in the presence of a magnetic field. 
 
 The five dimensional theory is described by the action  
\beq \la{5dAction}
S_5 = { 1 \over 16 \pi G_5  } \int d^5x \sqrt{g} (R - 2\Lambda_5 )- { 1 \over 4 g_5^2 } \int d^5 x \sqrt{g} F_{\mu \nu} F^{\mu \nu} ~,~~~~~~~\Lambda \equiv - { 6 \over R_5^2} 
\eeq  
which contains two dimensionless parameters $R_5^3/G_5$ and $R_5/g_5^2$ determining the (inverse) couplings of gravity and the gauge field. 

\subsection{The emergence of a two dimensional CFT from the four dimensional one} \label{TdFd}

In this subsection argue that, in the presence of a magnetic field, the four dimensional theory becomes effectively a two dimensional CFT. 

When we start with free massless four dimensional charged fermions, the two dimensional CFT comes from the lowest Landau level, which gives rise to massless fermions propagating along the magnetic field lines. So in this case it is relatively easy to see. The central charge is then $c_2 = N_f q$.

We now start from a holographic theory, or an $AdS_5$ space, and we  study the effects of adding a magnetic flux at the boundary of $AdS_5$.  We will derive that, as we move away from the boundary, the geometry transitions from $AdS_5$ to $AdS_3 \times R^2$. The presence of this $AdS_3$ factor is interpreted as the two dimensional CFT of the previous discussion. This necessary solution was discussed in \cite{DHoker:2009mmn}, and we review it below.

 We write the metric and gauge field ansatz consistent with the symmetries 
\beq \label{5dansatz}
ds^2 = R_{5}^2 \left[ e^{2 \lambda} (-dt^2 + dx^2 ) + e^{ 2 \sigma} (dy_1^2 + dy_2^2) + d\rho^2 \right] ~,~~~~~~~~~ A = B y_1 dy_2 
\eeq
Inserting this into Einstein's equations we find 
\bea  
0 &=& - 6 + 3 l_B^{-4} e^{ -4 \sigma} + \lambda'^2 + 4 \lambda' \sigma' + \sigma'^2  , ~~~~~\lambda = \lambda(\rho) ~,~~~~~~\sigma = \sigma(\rho)  \cr 
 0 &= &4 - 4 l_B^{-4} e^{ - 4 \sigma} - 2 \lambda ' \sigma' - 2  \sigma'^2 - \sigma''  ~,~~~~~~~~~l_B^4 \equiv    { 3 g_5^2 R_5^2 \over 4 \pi G_{5 } B^2 } \label{FlowEq}
\eea
We now make the following observations.

For large $\rho$ we have a solution $\lambda = \sigma = \rho$, where we neglect the term proportional to $ l_B$. This is the asymptotically $AdS_5$ form of the geometry. 
The term involving $l_B $ becomes important when 
\beq 
e^{ - \sigma } \sim  l_B 
\eeq
This is can be viewed as the distance scale where we have the transition between the four dimensional behavior and the two dimensional one. This scale is proportional to $1/\sqrt{B}$,  as we expect from general arguments, since $B$ is the only dimensionful quantity in the four dimensional sense.

At longer length scales $\sigma$ is constant, $e^{ -\sigma } = l_B$ and   the geometry becomes that of $AdS_3 \times R^2$, with an $AdS_3$ radius which is of order of the $AdS_5$ radius.  More precisely, in this region $e^{ - \sigma } = l_B $ and $R_3^2 = R_5^2/3$. The full numerical solution  of \nref{FlowEq} is indicated in Figure \ref{Flows} in appendix \ref{app:Flow}. Note that in the IR, when $e^{-\sigma} = l_B$, the proper size of the magnetic field in the transverse space goes to a constant independent of the original 4d flux, $Be^{-2 \sigma }  \propto \sqrt{{ g^2 \over R_5 } { R_5^3 \over G_5 } } $, which is just 
the ratio of the five dimensional $U(1)$ vs gravitational couplings. 

An important quantity is the two dimensional central charge \cite{Brown:1986nw,Strominger:1997eq}
\beq \la{cRS}
 c_2 = { 3 R_3 \over 2 G_3} =  {\sqrt{3}  R_5^3 \over 2 G_5 } { \Ac \over l_B^2 }  = 
2 \pi^{3/2} \sqrt{ R_5^3 \over G_5 } \times { \sqrt{R_5}  \over g_5  } \times q ~,~~~~~~~~~~ q =  { \Ac B \over 2 \pi } 
\eeq
 
Note that $c_4 \propto R_5^3/G_5$    is  the  ``central charge'' of the four dimensional theory (not to be confused with the two dimensional central charge on the left hand side of \nref{cRS}). The second factor is the effective five dimensional gauge coupling at the scale of the radius of $AdS_5$. Both of them should be larger than one for the validity of the theory. 
  

 If we were to put the four dimensional theory on $R^{1,1} \times  S^2$ with a large magnetic flux on the $S^2$, then we expect a slightly different solution. However, in the regime that the magnetic flux is large enough, so that the length scale $l_B$ is much smaller  than the radius of the $S^2$, then we expect that we can neglect the curvature of $S^2$ in studying the flow to the $AdS_3 \times S^2$ geometry in the IR. Indeed we check that this is the case in Appendix \ref{app:Flow}. For our application, we have that $l_B/r_e \sim \sqrt{R_5/r_e } \ll 1$, so that this approximation works. The situation is fairly similar when we put the four dimensional theory on $AdS_2 \times S^2$.  
  
 \subsection{The Randall Sundrum II model that we consider } 
 \la{RSII} 
 
 We will now discuss the full model under consideration.

 We imagine that we have a five dimensional gravity theory with a five dimensional  $U(1)$ gauge field, as in \nref{5dAction}. We then imagine that we have a so called ``Planck brane'' at a finite distance into the UV direction of  $AdS_5$. Its tension is fine tuned so as to make the cosmological constant  effectively zero, compared to the distance scales we consider. 
 In other words, we have the $AdS_5$ metric 
 \beq 
 ds^2 =  e^{ 2 \tilde \rho/R_5 } ( -dt^2 + d \vec x^2 ) + d\tilde \rho^2  ~,~~~~\tilde \rho< 0 
 \eeq
 Then the four dimensional metric is the metric at $\rho=0$ where the Planck brane sits. We imagine that the Standard Model fields live on the Planck brane. The four dimensional Newton constant is
  \beq
  { 1 \over G_4} = \half { R_5 \over G_5 }    \label{Gfour}
  \eeq
 where we have set to zero a possible Einstein term we can add on the brane. 
 
 Similarly, the five dimensional gauge  field leads to a four dimensional gauge field with a coupling 
 \beq  \label{gfour}
 { 1 \over g_4^2}  = { R_5 \over g^2_5 } \log\left( { L_{IR} \over R_5 } \right) 
 \eeq 
 where the logarithm can be interpreted as the running of the four dimensional coupling constant due to the charged matter \cite{Kaloper:2000xa}. It is IR divergent, but this is not really a problem for us. When we think about the near horizon region of the black hole geometry there is a natural IR cutoff which is $r_e$. Moreover, the fact that the 4d theory flows to an essentially two dimensional theory at scale $l_B$ provides an even shorter IR cutoff. So the log just gives us an extra factor of order $\log(l_B/R_5)$. 
 
  We are interested in making a wormhole of a fixed size $r_e$ as traversable as possible. This will involve maximizing $c_2$, for a fixed $r_e$. This both makes $\ell$ shorter and the binding energy stronger. From \nref{cRS} we see that the last two factors are essentially the same as the ones that appear in $r_e$ in \nref{redef}. So we want to maximize the first factor in 
  $c_2$ which amounts to making $R_5^3/G_5$ as large as possible. This in turn implies that we want to make $R_5$ as large as possible. 
  
  According to \cite{Lee:2020zjt}, the experimental upper bound on $R_5$ is roughly about $50 \mu m = 5 \stimes 10^{-5} m$. So we pick this value for the estimates below. Note that \nref{Gfour} implies that $ G_5^{-1/3} \gg 1 \, {\rm TeV}$.  
  
  The arguments we had above lead to a solution for this model if 
 $r_e \gg R_5$, so that we can think of the $AdS_5$ region as a special four dimensional field theory.  We are not using a full blown AdS/CFT duality, we are only making the observation that the 5d problem reduces effectively to a problem we can analyze using the 4d intuition. In fact, we are not going to find the full 5d solution explicitly, we will only argue that it should exist based on solving it by patches using effective field theory reasoning. 
 As we explained in the previous section the four dimensional solution depends only on the four dimensional gauge coupling and $c_2$. 
 These are given in \nref{gfour} and \nref{cRS}. The five dimensional geometry has a boundary that is determined by the four dimensional wormhole geometry. As it extends into the fifth dimension, it locally has an $AdS_5$ geometry in the UV region which transitions into an $AdS_3 \times S^2$ geometry at a distance  $l_B$ \nref{FlowEq} which is 
\beq
\label{lBval}
l_B = \left[ { 3 } \log \l r_e/R_5 \r \right]^{1/4} \sqrt{r_e R_5} 
\eeq
where we used the four dimensional equations \nref{redef} that determined the magnitude of 
the magnetic field in the throat region. We also estimated $l_B \propto \sqrt{r_e R_5}$ inside the logarithm in \nref{gfour} with $L_{IR} \sim l_B$, which yielded an extra $1/2$.   

 This implies that $ l_B$ is some intermediate scale between $R_5$ and $r_e$, so that the  flow to $AdS_3$ is well described by the discussion in section 
 \ref{FlowAdS}. Notice that this flow to a locally $AdS_3$ region occurs both in the wormhole region as well as the region outside where there is some magnetic field. As the magnetic field is weaker outside we flow for longer and the $AdS_3$ region is reached at a distance scale which is larger in the four dimensional space. Outside, in the region between the two mouths, the magnetic field also depends on the angular coordinate on the $S^2$ that surrounds  one of the mouths. This implies that the flow distance also depends on this angular dimension and the five dimensional geometry will not be a direct product. 
 
\begin{figure}[!ht]
\includegraphics[width=17cm,height=7cm]{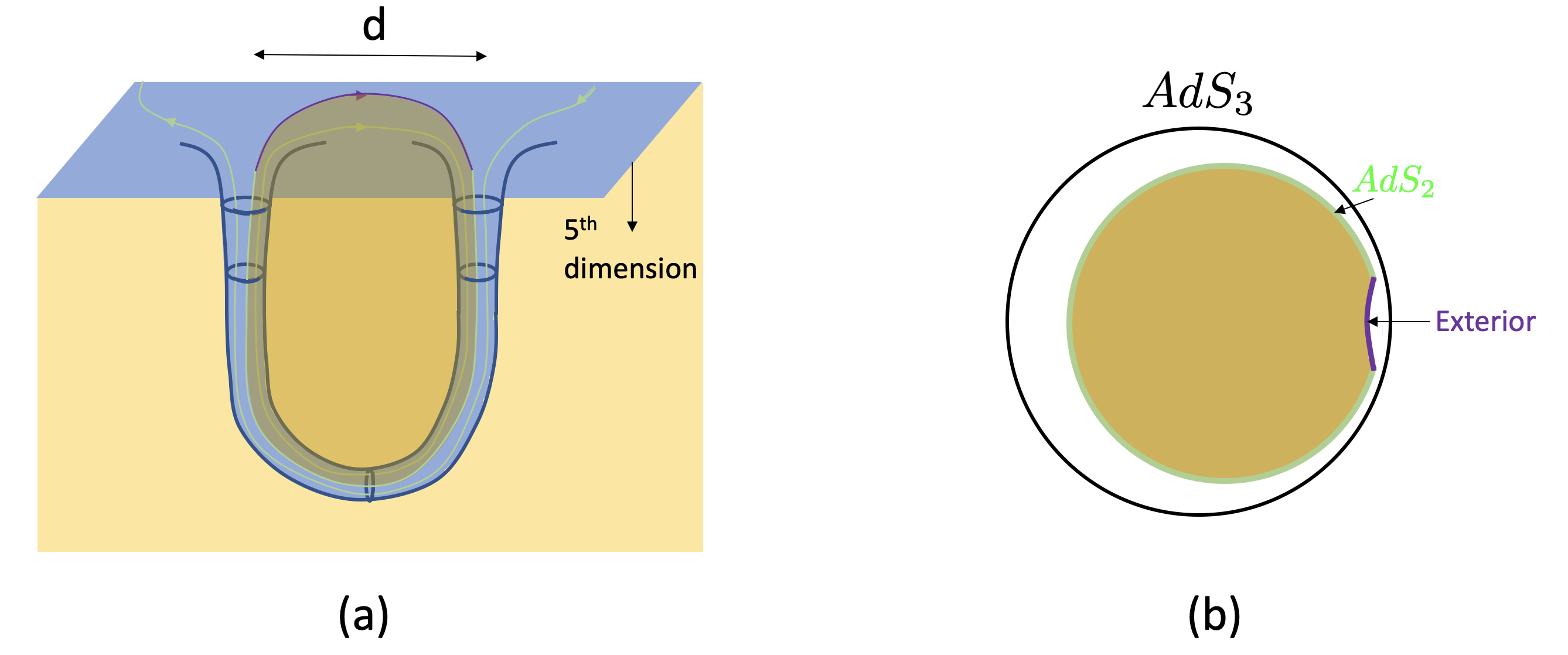}
\caption{ 
(a) We display the topology of the full five dimensional space. It is essentially filling the space ``outside'' the wormhole in the ambient space. We highlighted a special slice that ends on a particular magnetic field line.  (b) We display IR part of the geometry of the highlighted slice. This geometry is a portion of global $AdS_3$, the time direction is perpendicular to the page and we plot just the spatial direction.   The   Beyond the drawn region, the geometry goes over to the $AdS_5$ geometry as it approaches   the Planck brane. The   part of the magnetic field line that is outside the wormhole was colored 
purple.
For each magnetic field line, or each point on the $S^2$ we have a slice like this.   
 }
\label{Topology}
\end{figure}

 As we go deeper into the $AdS_3$ region, this space contains a time direction and the spatial directions will be as drawn in Figure \ref{Topology}(b). We have a circle which follows the magnetic field lines, where the $AdS_3$ goes over to the $AdS_5$ geometry. This circle is then ``filled in'' by the extra dimension, see Figure \ref{Topology}(b). This $AdS_3$ geometry leads to a classical negative gravitational energy corresponds to the negative Casimir energy in the 2d CFT interpretation. 
In this Figure the whole exterior region occupies a relatively short space, 
a portion equal  to $d/(\pi \ell)$, indicated in purple\footnote{The purple line in Figure \ref{Topology}(b) curves inwards because according to eq. (\ref{FlowEq}), 
distance $l_B$, where $AdS_3$ starts, is inversely proportional to the magnetic field. Since the magnetic field decreases as we go away from the wormhole mouths into the flat space region, we flow to a lower scale and thus  the metric scale factor is smaller. The green line in Figure \ref{Topology}(b) also approaches the boundary at the wormhole ends because the induced metric on this surface should be that of $AdS_2$.  
} in Figure \ref{Topology}(b).  
  In summary, 
  the full topology of the solution is the following. First, let us imagine fixing the two mouths of the wormholes. Since $g_{tt}$ never vanishes, we can just forget about the time direction and discuss only  the topology of the spatial directions.   
  The three spatial directions of the Planck brane have a non-trivial topology that roughly looks like that of Figure \ref{Topology}(a). If we compactify the point at spatial infinity, the  topology of these three spatial dimensions   is  $S^1 \times S^2$. After we add the extra dimension, the final topology of the spatial dimensions is essentially $D^2 \times S^2 $, where we fill in the $S^1$.

  The fact that the $S^1$ is contractible in the full geometry is interesting. The topological censorship theorems say that in classical general relativity we cannot have a non-trivial first homotopy group,  $\pi_1$.   This is allowed when quantum effects are included. However, what looks like a quantum Casimir energy in four dimensions is actually a classical effect in five dimensions (the negative classical energy of $AdS_3$).   So in five dimensions the topological censorship should work. It indeed works because the classical geometry in five dimensions has a trivial $\pi_1$. More physically, in four dimensions, the null ray that goes through the wormhole cannot be deformed continuously into one that stays  outside the wormhole. This is possible by moving the light ray into the fifth dimension. 
  
  We have argued that the solution exists, but one could imagine trying to find it numerically in a more explicit form.  
  
   As a simple example where we can discuss the geometry more explicitly, let us take the  four dimensional geometry to be the Einstein static universe, $R\times S^3$,  and we add the two wormhole mouths at antipodal points on the $S^3$.  This is not a solution of the four dimensional gravity theory we are considering. However, we can  sketch the form of the solution of  the five dimensional equations with these four dimensional boundary conditions.   The advantage is that  we have now an unbroken $SO(3)$ rotation symmetry and the geometry contains an $S^2$ factor (with varying radius) for each of the points of the rest of the three dimensions. The four dimensional geometry has two pieces. One is the wormhole interior with an $AdS_2 \times S_2$ geometry. The exterior region is $R \times S^3$. 
   Deep in the bulk the metric is $AdS_3$ in global coordinates. This continues up to a transition region, with roughly the shape indicated by the green and purple lines in Figure \ref{Topology}. 
After this region, the metric takes a locally $AdS_5$ form with a local geometry determined 
by the flow discussed in section \nref{TdFd}. It is possible to describe the metric
more explicitly and find the solution numerically, but we will not do it here.

 \subsection{Requirements for a humanly traversable wormhole } 
 
 Somebody traversing the wormhole will be subject to gravitational tidal forces, which are similar to the ones felt by somebody falling into  an extremal black hole. Since the curvatures are of order $1/r_e^2$, the tidal acceleration felt by the observer would be roughly $a \sim ({\rm size})/r_e^2$. We need to demand that this is smaller than the maximal sustainable accelerations which are about $20 g$  for short durations, where $g =9.8 \, m/s^2$  
  \beq \la{remin}
  20 g > a \sim { {\rm size }  \over r_e^2 }     ~~~~~~\to ~~~~~~~r_e  > 1.5 \stimes 10^7 m  \sim .05 s 
  \eeq
 where we assumed that the size is about $0.5 \, m$. We see that the size of the wormhole mouth should be very large, since we are very fragile. We have also written it as a time, since this sets proper time it takes to traverse the wormhole, which is $ \pi r_e$. 
 It is useful to express all the wormhole quantities in terms of $r_e$ 
 \beq \notag
 c_2 = 4 \pi { R_5 r_e \over G_4 \sqrt{\log(r_e/R_5)} } ~,~~~~~\ell = { 4 \over \pi } { r_e^2 \over R_5 }  \sqrt{\log(r_e/R_5)} ~,~~~~~~~E_{\rm bin} = - { \pi^2 \over 16} {R_5^2 \over G_4 r_e \log(r_e/R_5) } 
 \eeq
 Taking the minimal size for $r_e$ in \nref{remin}  and the 
largest possible value for $R_5$, of $50 \, \mu m$ \cite{Lee:2020zjt},  
we see that the parameters are 
 \beq \la{Params}
 r_e \sim  .05 \, {\rm s} ~,~~~~ c_2 \sim 7 \stimes 10^{72} ~,~~~~~ \ell 
\sim 3 \stimes 10^3 ~{\rm ly} ~,~~~~
 \gamma = { \ell \over r_e } \sim 2 \stimes 10^{12} ~,~~~~~~E_{\rm bin}  
\sim - 5  \stimes 10^9\, {\rm kg} 
 \eeq 
 where $\ell $ is given in light years.  We see that the traversal time as seen from the outside is rather long.

This time is basically unchanged as long as the distance, $d$, between the two 
mouths is smaller $d \lesssim \ell$. For $d$ of order $\ell$ the rotation 
frequency is much smaller than $1/\ell$. So if these wormholes were in an empty, 
zero temperature, flat space, they could exist for a while and be traversable. 

One might be worried that a highly boosted observer going through a weakly curved geometry will 
feel large tidal forces. This happens in general, but $AdS_2$ is boost invariant, so
the traveller does not feel anything special. However, we also need to take
into account that there could be small non-boost invariant perturbations to the
metric\footnote{We thank E. Martinec for prompting us to look at this issue.}.
These perturbations can be of order one near the mouth region, but as we go to
the center of the wormhole they are suppressed by a power of the ratio of the
redshift factor at the center. This works out to be of order $1/\gamma$ or
smaller \nref{Boost} (it is smaller for higher mass Kaluza Klein modes). Here
$\gamma$ is also the boost factor \nref{Boost}.  A boosted observer would see
tidal forces enhanced by a factor of $\gamma^2$. However, the duration of the
force is also decreased by a factor of $1/\gamma$, so that it integrates to a
total momentum transfer of order one.  This implies that the net size of the
effect of the perturbations is the same as it is in the mouth region, where
tidal effects are as in  \nref{remin}.

  \subsection{Some practical problems } 
   
  When an object  falls into the wormhole it develops  a very large boost factor, $\gamma$,  as seen by an observer at the center (or bottom) of the wormhole \nref{Params}.   In particular, if a CMB photon falls into the wormhole its energy will be enhanced by this factor. In addition, the wormhole traveller that finds it inside will see it with the square of this factor. So one would have to put the huge black hole inside a refrigerator in order to prevent this. Therefore, it seems exceedingly difficult to prevent this problem. 
   Of course, any particle that falls into the wormhole would create a related problem. So the wormhole has to be very clean before it can be traversed.
  
  If particles that fall into the wormhole scatter and lose energy then they would accumulate inside, contributing some positive energy that would eventually make the wormhole collapse into a black hole.    

This solution is assuming that the dark sector is
at a extremely low temperatures, smaller than $1/\ell$, or $10^{-26} eV$.
In our universe, even if we somehow arranged the cosmological evolution
to produce such low temperatures,  there would be processes that send energy
into the dark sector that heat it up to a much larger temperature. One simple
example of such a process is the soft radiation produced by particle collisions
in a star. This can have a 5$^{th}$ dimensional component, and the radiation
thus produced is larger than the purely four dimensional computation
\cite{weinberg1965infrared}, which by itself is enough to heat it to a much
larger temperature.


Another problem seems to be producing the wormhole in the first place. It would be interesting to understand whether they can be produced in the RS model. Since they require topology change,  this seems difficult.

 We highlighted the wormholes that are good enough to send a person. However, we could consider smaller wormholes, say of size $r_e \sim R_5$ which would correspond to a mass $M \sim 10^{23} \, kg$.   In this case, we can separate them by a large distance $d$, say over the solar system and then the traversal time would be comparable to their distance.  Of course, it is very small and the tidal forces would be huge. But we could use them to send very secret signals or qubits.

  \section{Conclusions } 
  
  We have argued that the Randall Sundrum II model
  allows for traversable wormhole solutions. In fact, it allows for solutions where the wormholes are big enough that a person could traverse them and survive. 
  
  From the outside they resemble intermediate mass charged black holes. Their big size comes from demanding that a human traveller can survive the tidal forces. They take a very short proper time to traverse, but a long time as seen from the outside. The traveller acquires a very large boost factor, $\gamma $, as it goes through the center of the wormhole. 
    
 We have argued this most clearly for the case that the wormhole exists in a
cold and flat ambient space, much colder than the present universe.  We
have {\it not} given any plausible mechanism for their formation. We have only
argued that they are configurations allowed by the equations.

  \textbf{Acknowledgments}

 We are grateful to A.~Almheiri, E.~Martinec, F.~Popov, L.~Randall and N.~Sukhov for discussions.

J.M. is supported in part by U.S. Department of Energy grant DE-SC0009988, the Simons Foundation grant 385600.

A.M. also would like to thank C.~King for moral support.
\appendix 

\section{More details on the flow} 
\la{app:Flow}

\begin{figure}[!ht]
\centering
\includegraphics[scale=1.2]{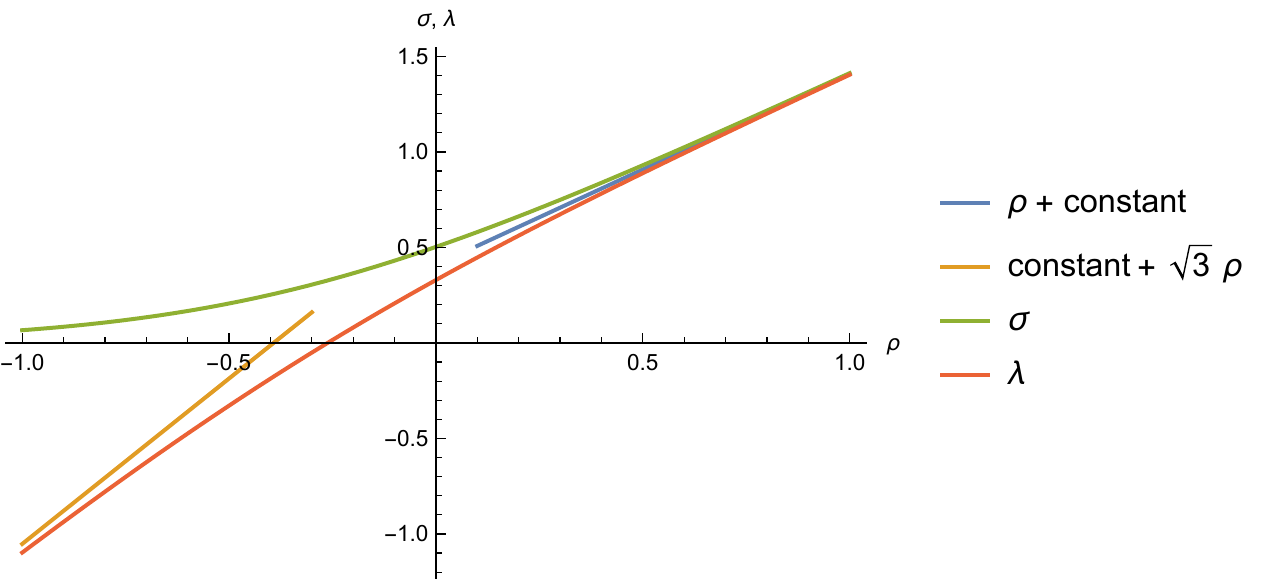}
 \caption{We show the numerical solution describing the flow. We have $AdS_5$ at large $\rho$, where $\sigma \sim \lambda \sim \rho + $constant. For negative  $\rho$ this goes to $\sigma \sim 0$ and $\lambda' = \sqrt{3}$. Besides plotting $\sigma$ and $\lambda$ we plotted the approximations that are valid when $|\rho|$ is large. 
 }
\label{Flows}
\end{figure}

In this Appendix we can include a numerical discussion of equations \nref{FlowEq}. Let us write them again after 
setting $l_B =1$ by a rescaling of $\sigma$. 
\bea \label{FIeq}
0 &=& - 6 + 3   e^{ -4 \sigma} + \lambda'^2 + 4 \lambda' \sigma' + \sigma'^2   \\  
 0 &= &4 - 4   e^{ - 4 \sigma} - 2 \lambda ' \sigma' - 2  \sigma'^2 - \sigma''   
\eea
These equations have only three integration constants. Two of them are trivial, one corresponds to shifting $\lambda$ by a constant and the other to shifting $\rho$ by a constant. We shifted $\sigma$ to simplify the equation. And we are left with only one non-trivial integration constant.  
We can solve for $\lambda'$ from the first equation \nref{FIeq} and replace it in the second. This gives an ordinary differential equation for $\sigma$. We solve it as follows. First we
note that $\sigma =0 $ , $\lambda = \sqrt{3} \rho + $constant is a solution. 
Then we expand in small fluctuations around this solution, assuming $\sigma$ is very small.   This linear equation has two solutions which are simple exponentials, one increasing and the other decreasing as a function of $\rho$. We want to impose the boundary condition that we have $AdS_3$ in the IR. This selects the increasing solution and also fixes the non-trivial undetermined integration constant for the equations. We then numerically solve the equations by choosing initial conditions at place where $\sigma$ is very small and given by the exponentially increasing solution. 
This leads to the numerical solutions in Figure \ref{Flows}. 
 
We now consider the case that we start from the theory on a two sphere. In other words we have $R^2 \times S^2$ at the boundary with $q$ units of magnetic flux, or $F = { q \over 2} \sin \theta d\theta d\varphi$. In this case the metric ansatz is the same as \nref{5dansatz} replacing $dy_1^2 + dy_2^2  \to d\theta^2 + \sin^2 \theta d\phi^2 $. By conformal symmetry we set the radius of the two sphere to one. 
The new equations are now 
\bea \label{FlowCur}
0 &=& - 6 -  e^{ - 2 \sigma } + 3 \hat l_B^{-4} e^{ -4 \sigma} + \lambda'^2 + 4 \lambda' \sigma' + \sigma'^2   \cr 
 0 &= &4 + e^{ - 2 \sigma } - 4 \hat l_B^{-4} e^{ - 4 \sigma} - 2 \lambda ' \sigma' - 2  \sigma'^2 - \sigma''  ~,~~~~~~~~~\hat l_B^4 \equiv    { 3 g^2 R_5^2 \over  \pi G_{5 } q^2 } \label{FlowEqa}
\eea
We see that the curvature of the sphere gives rise to the extra term involving 
$e^{-2 \sigma}$. Now the quantity $\hat \ell_B$ is a distance in units of the radius of $S^2$. The radius of the $S^2$ in the IR is given by solving the second equation with $\sigma' =0$ which gives 
\beq
 e^{ - \sigma } =  \hat l_B +  { ({ \hat l_B})^3 \over 16 } + \cdots ~,~~~~~{\rm for } ~~~\hat l_B \ll 1 
 \eeq 
 The second term is the correction due to the curvature of the sphere. 
The radius of $AdS_3$ has a similar small correction. 
   We are indeed interested in the case that $\hat \ell_B = { l_B \over r_e} \propto \sqrt{ R_5 \over r_e }  \ll 1$, see equation \nref{lBval}. 
   We can also numerically solve \nref{FlowCur} and obtain answers similar to those in Figure \ref{Flows}.

  \section{Rotation and the emission of radiation } 
  \label{Rotation} 
  
  Here we briefly review the discussion in \cite{Maldacena:2018gjk} for (temporarily) stabilizing the wormhole mouths by rotation. 
  When the mouths are far away, $d\gg r_e$ we can use the Kepler formula 
  \beq 
  \Omega = \sqrt{ r_e \over d^3 }
  \eeq
  We would like to be in a regime where this is much smaller than the energy gap 
  $1/\ell$ of the wormhole. This would ensure that the throat is unaffected by the radiation and 
Unruh quanta \cite{Letaw:1979wy}. For the parameters in \nref{Params} we get the condition $d > .5 $ ly.  If the two black holes were orbiting the galaxy we also get that the frequency would be much smaller than $1/\ell$.

  We now consider the emission of four dimensional gravity waves as well as four dimensional ``dark'' $U(1)$ radiation. These rates are given by 
\beq
\l { d E \over d t }\r_{\rm grav} = { 2 \over 15 }{ r_e^2 d^4 \Omega^6 \over G_4 }   
 ~,~~~~~~~~~~~ \l {d E \over dt } \r_{\rm U(1) }= \frac{ \pi}{ g_4^2} { 2 \over 3 }  q^2 d^2 \Omega^4  = { 2 \over 3} { r_e^2 d^2 \Omega^4}
\eeq
We see that the $U(1)$ radiation dominates over the gravitational one, since for us $d\Omega < \ell \Omega \ll 1$. 

 The time, $T_{\rm lifetime}$, that it takes the black holes to collapse on each other is
\beq
T_{\rm lifetime} \propto \frac{m v_g^2}{2\l d E/ dt  \r_{\rm U(1) }} = { 3 \over 4 } { d^3 \over r_e^2 }  \eeq 
 For the parameters in \nref{Params}, and for the lowest $d \sim .5$ ly, this gives a lifetime of about $10^{16}$ years.

\mciteSetMidEndSepPunct{}{\ifmciteBstWouldAddEndPunct.\else\fi}{\relax}

\bibliographystyle{utphys}
\bibliography{ref}{}

\end{document}